\documentclass[preprints,article,accept,moreauthors,pdftex, hyperref]{definitions/mdpi} 

\newcommand{\tgate}{t_{gate}}
\newcommand{\tdelay}{t_{delay}}

\firstpage{1} 
\pubvolume{xx}
\issuenum{1}
\articlenumber{5}
\pubyear{2019}
\copyrightyear{2019}
\history{Received: date; Accepted: date; Published: date}

\newcommand{\kk}{\mathbf{k}}
\newcommand{\qq}{\mathbf{q}}

\Title{Theory of time-resolved optical conductivity of superconductors:
comparing two methods for its evaluation}

\Author{John P. Revelle $^{1}$, Ankit Kumar $^{1}$\orcidB{} and Alexander F. Kemper$^{1,}$*\orcidA{}}

\AuthorNames{John P. Revelle, Ankit Kumar and Alexander Kemper}

\address{%
$^{1}$ \quad Department of Physics, NC. State University
}

\corres{Correspondence: akemper@ncsu.edu}

\abstract{
Time-resolved optical conductivity is an oft-used tool
to interrogate quantum materials driven out of equilibrium.
Theoretically calculating this observable is a complex 
topic with several approaches discussed in the literature.
Using a non-equilibrium Keldysh formalism and a functional
derivative approach to the conductivity, we present a 
comparison of two particular approaches to the calculation
of the optical conductivity, and their distinguishing 
features, as applied to a pumped superconductor.
The two methods are distinguished by the relative
motion of the probe and gate times; either the probe
or gate time is kept fixed while the other is swept.
We find that both the methods result in same qualitative features 
of the time-resolved conductivity after pump is over.
However, calculating the conductivity by keeping the gate fixed removes 
artifacts inherent to the other method.
We provide software that, based on data for the first method, is
able to construct the second approach.}


\begin{document}

\section{Introduction}
Time-resolved optical conductivity is one
of the workhorse experiments for studying quantum materials driven out of equilibrium. Recent advances in THz
technology have enabled the time-resolved measurement
of the conductivity at low frequencies,  and this approach has
been applied to a variety of systems, including
superconductors driven out of equilibrium, where several
novel features have been observed.  These include a low-frequency
upturn in the inductive response \cite{2016Mitrano}, which indicates a potential
enhancement of superconductivity and oscillations at a frequency
of twice the superconducting gap ($2\Delta$) that has been attributed
to the Higgs amplitude mode of the superconductor \cite{2013Matsunaga, 2014Matsunaga,2016Murakami, 2016Tsuji,  2017Kennes, 2019Silaev,
2019Murotani, 2019Shimano, 2019Jujo}, although the contribution from light-induced excitation of the Cooper pairs is 
also shown to be important \cite{2015Cea, 2016Cea, 2019Udina}.

From the theoretical side, the calculation of time-resolved optical conductivity has been limited to few cases,
or evaluated
\cite{
2014Krull, 2015Orenstein, 2015Cea, 2016Shao, 2017Kennes} using simple models for the electronic states and the time evolution. Notable exceptions
are Eckstein et al.\cite{eckstein2008theory}, Tsuji et al.\cite{2009Tsuji} and Kumar et al.\cite{kumar2019higgs}
who used a non-equilibrium Green's function approach
for the driven electronic states and in one case a numerical
functional derivative approach to calculate the
optical conductivity. The important advance of the latter
is that it includes the vertex corrections due to the
included interactions automatically.
Kumar et al.\cite{kumar2019higgs} studied the time-resolved optical conductivity
of a driven superconductor, and observed signatures
of the Higgs oscillations across the spectrum.

\begin{figure}[htpb]
    \centering
    \includegraphics[width=0.5\textwidth]{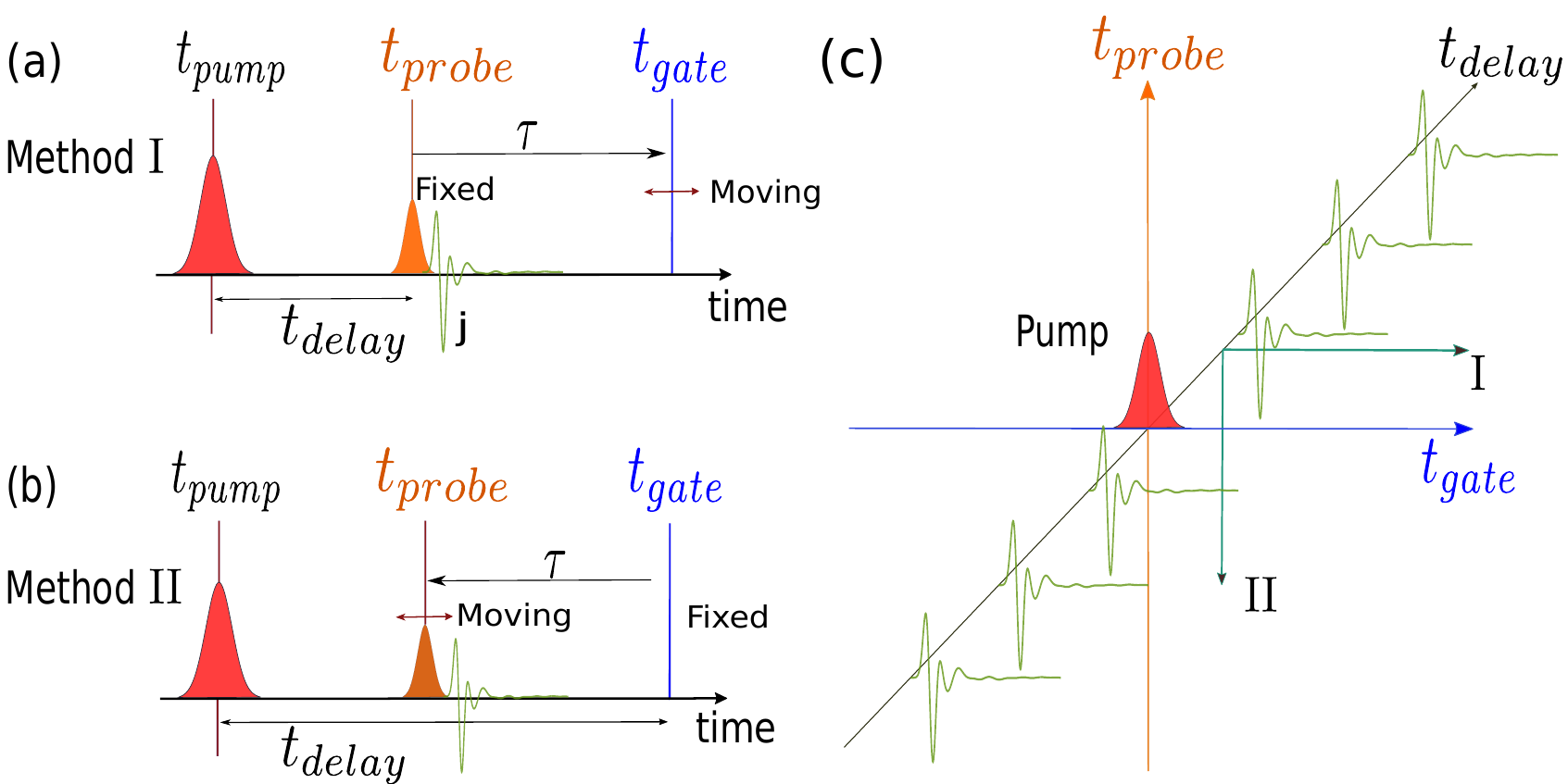}
    \includegraphics[width=0.43\textwidth]{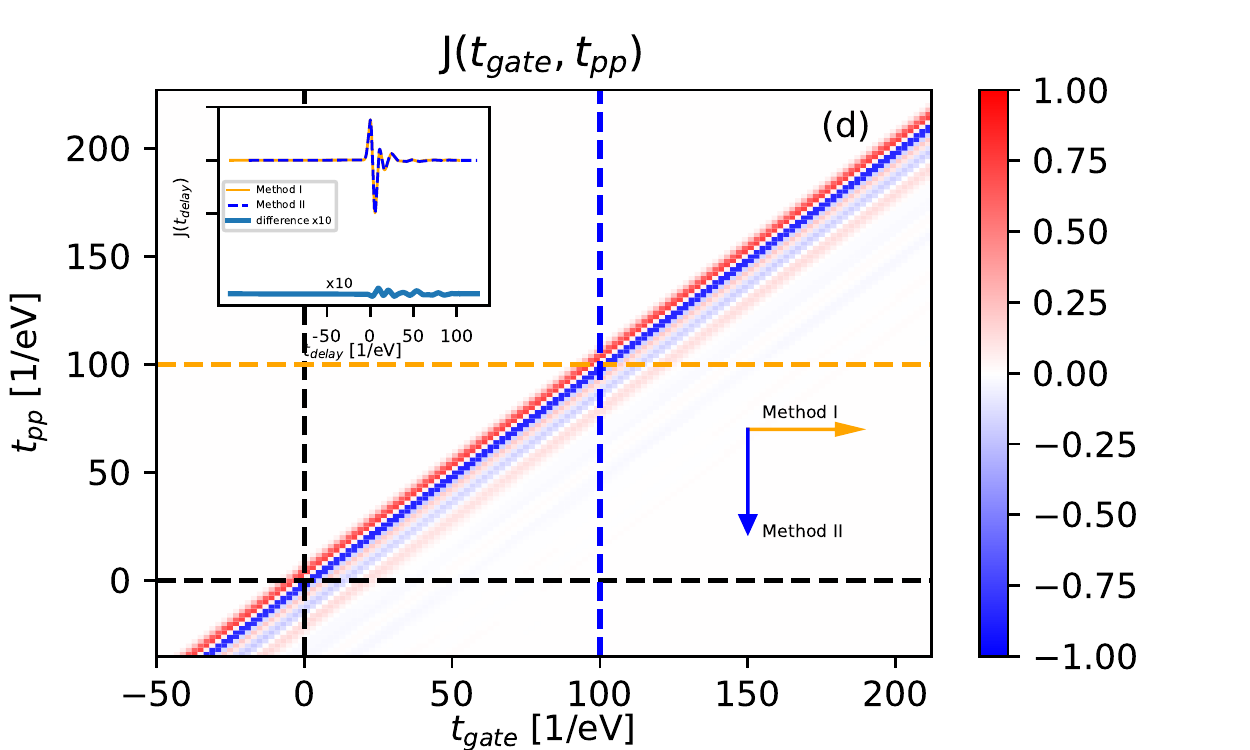}
   \caption{(Color online) Two Methods for computing the
    optical conductivity. The three time points, $t_{pump}$, $t_{probe}$ and $\tgate$ are shown, as well as the variable
    time to be Fourier transformed over: $\tau$. The green curve shows a sample probe current $(J(t))$ obtained with the
    non-equilibrium Keldysh method. The pump time defines $t=0$. (a) Method I sweeps $\tgate$ to later times and the
    delay-time $\tdelay$ is set by the pump-probe spacing. (b) Method II sweeps the probe pulse at$t_{probe}$ to
    earlier times and $\tdelay$ is set by the pump-gate spacing.  (c) Schematic of the observed currents in the
    $(\tgate$, $t_{probe})$ plane, with the relative time points from (a) and (b) indicated.  Methods I and II
    correspond to horizontal/vertical cuts in this plane, respectively.  (d) Probe currents obtained from the
    non-equilibrium Keldysh formalism after interpolation. Inset: horizontal and vertical cuts of the current at
    $\tgate=t_{pp}=100$. Beneath is a plot of the difference (Method II - Method I) between the two curves, scaled by a factor of 10 to increase visibility. 
    }
    \label{fig:overview}
\end{figure}

Fundamentally, the conductivity $\sigma$
is the linear proportionality between
the applied electric field $E(t)$ and the resulting current $J(t)$. In the
time domain, this is expressed as
\begin{align}
    J(t) = \int_{-\infty}^{t} d \bar{t} \sigma(t-\bar t) E(\bar t),
\end{align}
where we have suppressed vector indices for clarity. That is,
we apply a field at some time, and observe the resulting current at some
time later.
Given
this relation, we may take a functional derivative to obtain
the conductivity
\begin{align}
    \sigma(t-t') = \frac{\delta J(t)}{\delta E(t')},
\end{align}
or equivalently, a ratio in the frequency domain
\begin{align}
    \sigma(\omega) = \frac{J(\omega)}{E(\omega)}.
\end{align}
Out of equilibrium, the situation becomes more complex.  There are now
three separate time points: the pump time, the probe time, and the time
at which the current is measured (the gate time). These three are
illustrated in Fig.~\ref{fig:overview}.
The presence of the
pump pulse which induces system dynamics breaks the time-translation invariance,
which implies the equation for the conductivity now reads
\begin{align}
    \sigma(t,t') = \frac{\delta J(t)}{\delta E(t')},
\end{align}
and an ambiguity arises for the evaluation in the frequency
domain. That is, given that there
are now three fixed points in time rather than two, which temporal axis should be Fourier transformed? And, which ones (or which differences)
correspond to $t$ and $t'$?

The measurement or calculation of the optical conductivity
is typically performed with a pump and a probe at
times $t_{pump}$ and $t_{probe}$, respectively. 
The simplest approach is to measure the emitted field
after the probe as a function of sampling time (by a {\it gate}) 
and take a Fourier
transform along this axis, using the pump-probe separation
$t_{probe}-t_{pump}$ as the time delay axis in $\sigma(\omega, \tdelay)$.
This is schematically shown as ``Method I'' in Fig.~\ref{fig:overview}(a).
However, while the signal is collected, the dynamics
induced by the pump are still occurring in the
system, which are in a sense averaged over the time
during which the emitted field is measured. To remedy this,
another approach is used where the pump and gate
time
are kept at a fixed separation, and the probe pulse is swept 
backwards\textemdash this is shown as ``Method II'' in the figure. 
This second method has the advantage that the
system is always in the same state after the pump
whenever the measurement occurs (the probe is assumed
to be small and to not affect the dynamics). As was pointed
out here \cite{1999Kindt, 2002Nemec, coslovich2017ultrafast}, this remedies
issues such as the appearance of dynamics before the pump
occurs (termed ``perturbed free induction''). In this work,
we will apply both Methods to obtain the conductivity
of a pumped superconductor, and contrast the
approaches. 

\section{Methods}

The conductivities, regardless of which Method, are determined
by a functional derivative of the current $J(\tau)$.  The current is 
obtained by a non-equilibrium Keldysh Green's function formalism:
the self-consistent solution of the Dyson equation on the
Keldysh contour \cite{kemper2014effect}.
The equations of motion are solved in the superconducting
state using a the Nambu Green's functions\cite{2015Kemper},
with strong-electron phonon interactions mediating
the pairing interactions. In addition to electron-phonon
interactions (which also scatter in addition to providing
the pairing glue), we include impurity scattering
to consider the dirty limit of BCS and its resulting
signal below the energy of the pairing boson \cite{1958Mattis, 1991Zimmermann}.
The parameters for the calculation are listed in Tab.~\ref{tab:params}.
 These parameters were chosen for simplicity of calculation and do not represent any specific material; they may be adjusted to simulate real materials. 
Samples of the resulting currents are shown in Fig.~\ref{fig:overview}.

We calculated the conductivity using two Methods, as illustrated in Fig.~\ref{fig:overview}, and as explained here. In both cases, a current is measured as a function of a time
$\tau$, and a functional derivative is performed numerically by Fourier transforming
$J(\tau)$ and the electric field, and taking the ratio. The difference arises
in which time is kept fixed, and which is swept to evaluate the current $J(\tau)$.
There are three time points. $t_{pump}$ is the arrival time of the pump pulse, 
which is used as time zero. $t_{probe}$ is when the probe pulse hits the sample.
Finally, $\tgate$ is the time when the generated current is measured. The relative
time between the pump and the probe is $t_{pp} \equiv t_{pump} - t_{probe}$.

{\it Method I}: we compute for fixed values of $t_{pp}$ the ratio $\sigma(\omega,\tdelay) = \frac{\text{J}(\omega,t_{pp})}{\text{E}(\omega,t_{pp})}$, taking the Fourier transform along the $\tgate$ axis (in the horizontal direction in Fig.~\ref{fig:overview}(d)). This Method is
from a computational perspective straightforward since it simply involves the application
of two pulses, and calculating the resulting current.

{\it Method II:} $\tgate$ is kept at a fixed distance from the pump, and the probe is swept backwards to generate the
current $J(\tau)$. Then, for fixed values of $\tgate$ we compute the ratio
$\sigma(\omega,\tdelay) = \frac{\text{J}(\omega,\tgate)}{\text{E}(\omega,\tgate)}$, taking the Fourier transform
along the $\tau$ axis [in the vertical direction in Fig.~\ref{fig:overview}(d)]. This Method is computationally
more complex since a large number of pump-probe delay sets need to be generated. Here, we have taken data
generated from method I and performed Akima spline interpolation\cite{akima1970new} to be able to take vertical
cuts. The interpolation results are shown in Fig.~\ref{fig:overview}(d).

The advantage of Method II is that the system dynamics, which are driven by the pump, are always in the same
state when the current is measured (at $\tgate$). Since the goal of the measurement is to determine
the pump-induced dynamics, this Method may be able to more selectively observe these and provide 
better time resolution. In Method I, the effective time resolution for the dynamics is set by 
the decay time of the current (signal-length), which may be long and is generally not known in advance.
This long decay time (and thus long effective resolution) produces an averaging over the system dynamics,
which may obscure them or in extreme cases hide them if the pump-induced dynamics is comparable to the
signal-length. 

From an experimental viewpoint, there appears to be a preference
towards Method I, potentially due to its simplicity and the long relaxation of the excitations compared to the probe width (see e.g.
\cite{2014Matsunaga, 2018aYang, 2016Mitrano}. A notable
exception is Ref.~\cite{coslovich2017ultrafast}, which
discusses the perturbed free induction decay in some detail.
As we will demonstrate below, Method II has potential
upsides which may prove useful within the experimental
context.

\section{Results}

To demonstrate the difference between the two approaches,
we consider a driven system that has interesting dynamics
after the pump\textemdash a pumped superconductor. This system
shows non-trivial changes in the conductivity, most notably
oscillations of the superconducting order. Oscillations
are complex when it comes to evaluating them in the
optical conductivity since this requires a Fourier transform;
in principle this could simply average over the oscillations
and result in a peak in the conductivity rather than any
time-dependent behavior, and thus we expect the two methods
to show marked differences here.

\begin{figure}[htpb]
        \centering
        \includegraphics[width=0.98\textwidth, clip=true,trim= 60 120 140 120, ]{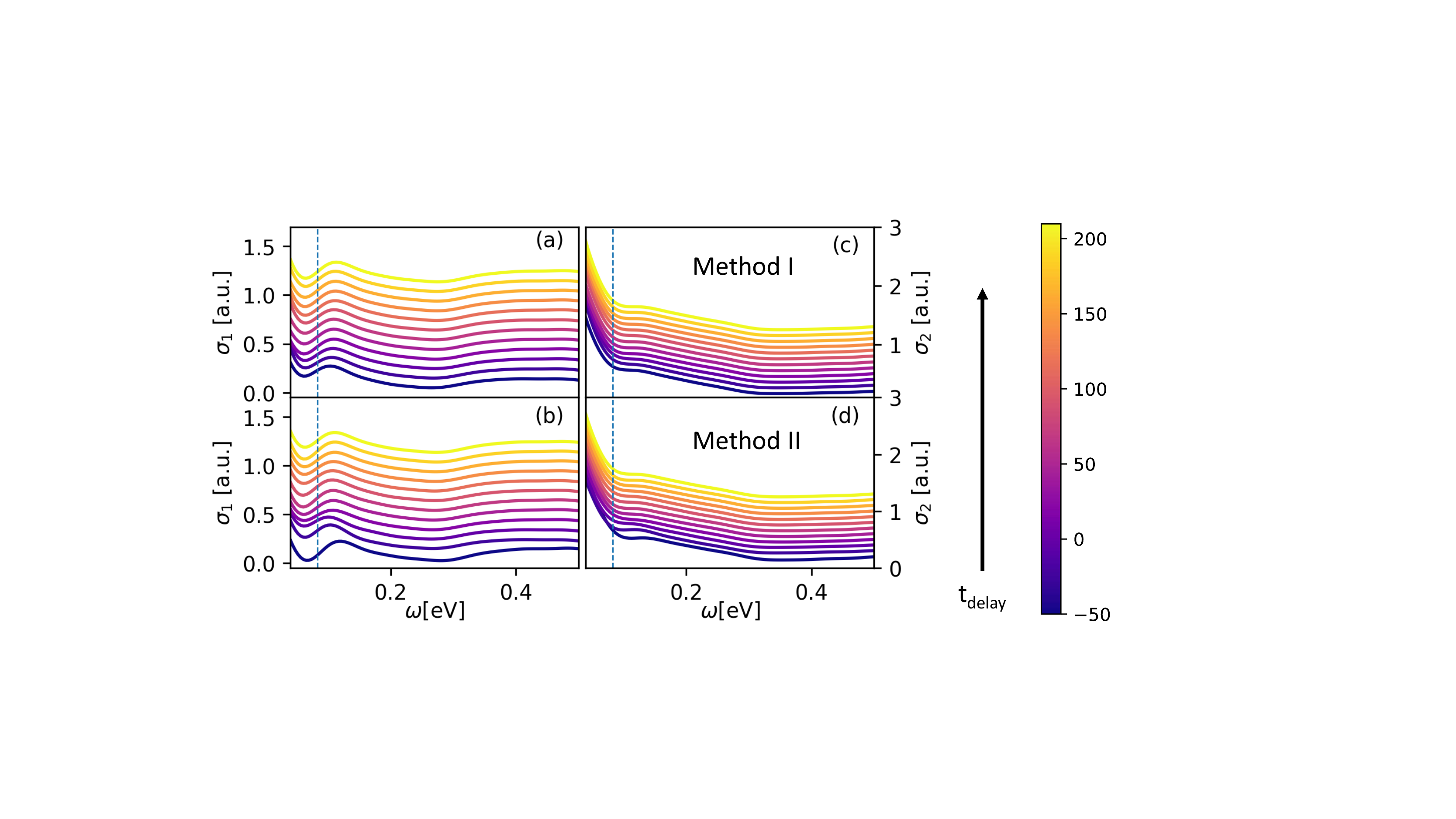}
        \caption{
            (Color online) The real $(\sigma_1)$ and imaginary $(\sigma_2)$ parts of the conductivity obtained with both Method I and Method II as a function of their respective $\tdelay$. (a)/(b) $\sigma_1$ obtained via Method I/II. (c)/(d) $\sigma_2$ obtained via Method I/II. In each panel the dashed line corresponds to the frequency $2\Delta_{eq} = $ 0.083 eV. To increase visibility of each curve there is a fixed offset between each conductivity. A video is included as a supplement.
            } 
        \label{fig:sigmas}
\end{figure}

Fig.~\ref{fig:sigmas} displays the evolution of the real  $(\sigma_1)$ and imaginary $(\sigma_2)$ parts of the conductivity computed with both Methods as a function
of their respective $\tdelay$ values. 
The bottom curves corresponding to $\tdelay = -50.0$ indicate the conductivity components in the equilibrium state, before the pump is applied.
The conductivity shows the expected features for a strong-coupling BCS superconductor in the presence of impurity scattering: in $\sigma_1$ an upturn at low frequencies in the real part, a
step near $2\Delta$, and a minimum at $\Omega + 2\Delta$, and in $\sigma_2$ a divergence
towards low frequencies.  As the pump is applied, the superconductor is partially
melted and the features who positions involve $\Delta$ red-shift.

In addition to a reduction in the order parameter,
the system exhibits Higgs (or Anderson-Higgs) oscillations, which are an oscillatory decay in the relaxation of the excited population of the Cooper pairs in superconducting condensates subject to perturbation by ultrafast pump fields\textemdash these were previously discussed based on similar calculations using the non-equilibrium Keldysh formalism\cite{2015Kemper,2016Tsuji,kemper2017review,kumar2019higgs}. Higgs oscillations arise here due to a time-dependence of the superconducting order parameter and we observe them in the time-dependent conductivity $\sigma(\omega,\tdelay)$ as time-dependent oscillations
of the gap edge and minimum around the phonon energy.
It is important to note that a critical aspect of the method for
observing the Higgs oscillations with Method I is that the 
probe current decays. If this were not the case, Method I would
effectively have no time resolution, and only oscillations in
the peak height would be visible \cite{2014Krull}.

\begin{figure}[htpb]
        \centering

        \includegraphics[width=0.99\textwidth, trim=0.05cm 0.02cm 0.02cm 0.05cm, clip=true]{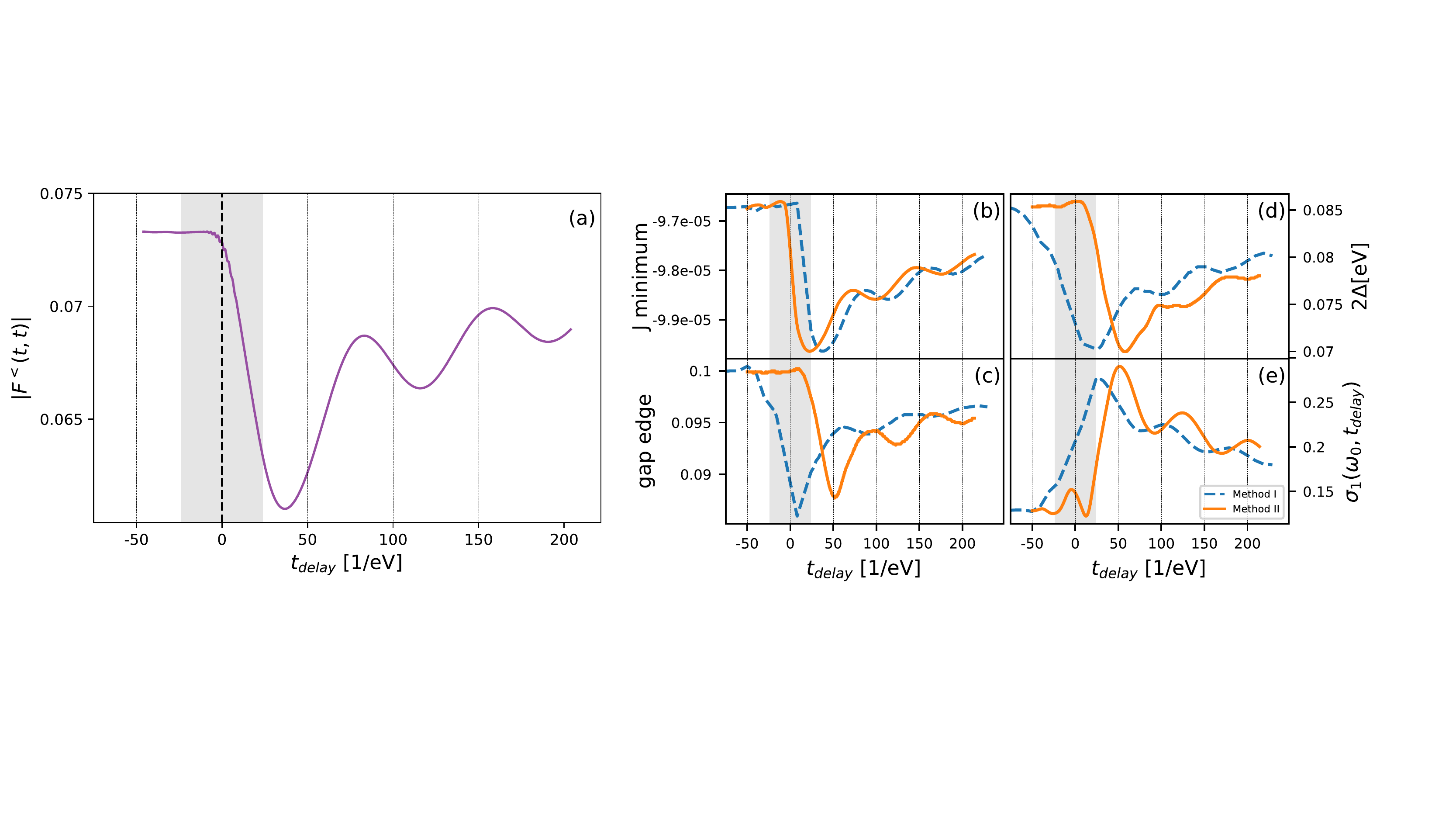}
        \caption{
            (Color online) Comparison of the two Methods
            applied to the dynamics after a pump. The region where the pump
            is active is indicated by a shaded region around $\tdelay=0$.
            (a) The equal-time anomalous Green's function.
            (b) Minimum probe current. 
            (c) Gap edge determined as discussed in the text
            (d) Gap-induced shift in the location of the minimum around the phonon energy in $\sigma_1(\omega, \tdelay)$.
            (e) $\sigma_1(\omega_0,\tdelay)$ at a fixed frequency around the coherence peak $\omega_0\approx 2\Delta_{eq}$.
            }   
        \label{fig:4panel}
\end{figure}

Following the analysis of Kumar et al.\cite{kumar2019higgs}, we further investigate the time-evolution of the superconducting order parameter in the aftermath of the pump
by considering the dynamics of four time-resolved quantities as functions of $\tdelay$ for the two Methods:
\begin{enumerate}
    \item The probe current minimum, which was demonstrated to be a measure of
    the order parameter\cite{2013Matsunaga,kumar2019higgs}.
    \item The location of the gap edge in $\sigma_1 (\omega,\tdelay)$, which we define to the the point $\omega_{edge}$ on the frequency axis where the mean of $\big( \sigma_1^{sc}/\sigma_1^{ns} \big)_{max}$ and  $\big( \sigma_1^{sc}/\sigma_1^{ns} \big)_{min}$ is located within
    the range from $\omega=0$ to $\omega \approx 2\Delta_{equilibrium}$. 
    \item The location of the $\sigma_1 (\omega,\tdelay)$ minimum about the phonon frequency $\Omega$ (measured with respect to $\Omega$).
    %
    \item The conductivity at a fixed frequency: $\sigma_1(\omega = 0.083,\tdelay)$. 
\end{enumerate}
These four quantities, obtained using both Methods, are shown in Fig.~\ref{fig:4panel}, panels (b)-(e), in the order in which they were discussed. The time axis $\tdelay$ is provided by $t_{pp}$ and $\tgate$ for Methods I and II, respectively. For reference, we also show the
anomalous ``density'': $F^<(t,t)$. This quantity is an instantaneous
measure of the superconductivity in the system; in equilibrium it
is equivalent to the right hand side of the gap equation, summed
over momenta:
\begin{align}
    F^<(t,t) = -i\sum_{\mathbf{k}} \frac{\Delta_{eq}}{2E_\mathbf{k}}
    \tanh\left(\frac{E_\mathbf{k}}{2T}\right),
\end{align}
where $E_\mathbf{k}=\sqrt{\xi_\mathbf{k}^2 + \Delta_{eq}^2}$, and
$T$ is the temperature.  Although this is an imperfect
measure of the amount of superconductivity in the system because
it only captures the amplitude of the order and
not the phase\cite{PhysRevB.99.241111}, it has shown to be
correlated with the spectral gap and its dynamics\cite{2015Kemper}.

There are several differences between the two Methods, which we will now discuss.
In all cases, the curves show a suppression of the gap, followed by the characteristic
Higgs oscillations, which were discussed in detail in Ref.~\cite{kumar2019higgs}.

However, the most striking difference is an apparent horizontal shift between the 
curves.
For the Fourier transformed quantities obtained from the conductivity 
(panels
(c)-(e)), the dynamics using Method I occur earlier by approximately 
50eV$^{-1}$.
These shifts are due to a mechanism termed ``perturbed free induction 
decay,'' where
for negative times the pump arrives while the probe-induced current
is still decaying, causing an earlier than expected observation of
changes in the conductivity due to the pump [see e.g. the third
current in Fig.~\ref{fig:overview}(c)].
This is in particular relevant for low-frequency features in the 
conductivity, which by nature require long time signals. This effect
produces both the earlier appearance of the maximal change and following
shift of the Higgs oscillations, but also the appearance of changes before
the pump is active. In contrast, Method II does not have this artifact
\textemdash the maximal change
occurs more closely to where $F^<(t,t)$ reaches its maximal change.

In the case of the minimum current, the reverse occurs: the current
from Method II reaches its largest change earlier than Method I.
Here, this is due to the delay in when the minimum current occurs: it
appears some fixed time after the pump, which in Method I shifts the time
to higher values.
In comparing the currents, Method II more correctly identifies when the change in the gap occurs\textemdash it stops decreasing when the pump is off.

\section{Summary}
We have evaluated the conductivity of a pumped superconductor with two
experimentally accessible methods. The methods involve two arrangements
of the three times involved in an optical pump-probe experiment:
the pump, probe, and gate times.
The conductivities obtained from the two Methods are qualitatively similar,
but differ in some key details. The pumped superconductor typically
exhibits a suppression, followed by recovery with Higgs oscillations.
Both methods discussed here exhibit these features, but with notable
differences.  First, the (simpler) Method I observes ``perturbed free
induction'' changes before the pump arrives. More generally,
due to an effective averaging over the current decay time, Method
I's features are somewhat smeared.  In contrast, Method II
has sharp changes, and exhibits no changes before the pump arrives.
Our results suggest that Method II offers improved time resolution over
Method I, although Method I does reproduce some of the observed
effects.


\section{Materials and Methods}

The simulations were performed with the self-consistent
non-equilibrium Keldysh formalism described in
Ref.~\cite{kemper2014effect}. The Holstein  Hamiltonian is used to simulate a phonon-mediated, $s$-wave superconductor
on 2D square lattice 
 \begin{align}
     \mathcal H &= \sum_{\kk,\sigma}\xi(\kk) c^\dagger_{\kk,\sigma} c^{\phantom\dagger}_{\kk, \sigma} + \Omega  \sum_{\qq}
                             \left(b_{\qq}^\dagger b^{\phantom\dagger}_{\qq} + \frac{1}{2} \right) 
               +\frac{g}{\sqrt{N}}\sum_{\substack{\sigma\\\kk,\qq}}c_{\kk+\qq, \sigma}^\dagger
         c^{\phantom\dagger}_{\kk,\sigma} \left( b^{\phantom\dagger}_{\qq} + b_{-\qq}^\dagger \right)   
         + \sum_{i,\sigma} V_i c_{i,\sigma}^\dagger c_{i, \sigma}
 \end{align}
 Here, $\xi(\kk)$ ($= -2 \mathrm{V_{nn}} \left[ \cos(k_x) + \cos(k_y) \right] -\mu$) is the nearest neighbor tight-binding
 energy dispersion with hopping parameter $V_{nn}$ measured relative to the chemical potential $\mu$, $c^\dagger_\kk, c^{\phantom\dagger}_\kk$ ($b^\dagger_\qq,
 b_\qq)$ are the standard creation and annihilation operators for an electron (phonon), $g$ is the momentum-independent
 $e$-ph coupling constant, and $\Omega$ is the frequency for the Einstein phonon. $V_i$ is the coupling between electrons
 and impurities which are distributed randomly on lattice sites. 

We used the parameters listed in Tab.~\ref{tab:params}.  An oscillating Gaussian pump pulse with a width $\sigma_p$ and
a central frequency $\omega_p$ was applied, followed by a probe pulse of similar shape but with $\sigma$ and $\omega$ as
width and central frequency, respectively. As illustrated in Fig.~\ref{fig:overview}, the pump-probe delay time was
varied.  The generated data was interpolated with Akima splines \cite{akima1970new} before taking Fourier transforms in
the two directions indicated in the figure.

The software and data used in this manuscript is publicly available at \cite{grendelgithub}.

\begin{table}[htpb]
    \centering
    \caption{Parameters used in the simulation}
    \setlength{\tabcolsep}{8pt}
    \begin{tabular}{lll}
        \hline
        \hline
        Phonon frequency $(\Omega)$ & 0.20 eV \\
        Phonon coupling $(g^2)$ & 0.12 eV  \\
        Impurity coupling ($\langle V_i\rangle^2 $) & 0.01 eV \\
        \hline
        Band parameters & $V_{nn} = 0.25$ eV, $\mu = 0.0$ eV \\
        Temperature & $\beta = 140 $ eV$^{-1}$\\
        Pump pulse & $\omega_p = 1.5$ eV, $ \sigma_p = 8 $ eV$^{-1}$ \\
        Probe pulse & $\omega = 0.01 $ eV , $\sigma = 3$ eV$^{-1}$ \\
        \hline 
        \hline

    \end{tabular}
    \label{tab:params}

\end{table}

\newpage


\authorcontributions{J.P.R. was responsible for the Methodology, software, and parts of the visualization. A.K. was responsible for the investigation (data collection) and parts of the visualization. A.F.K. was responsible for the conceptualization, funding acquisition, project administration, and writing.}

\funding{This material is based upon work supported by the National Science Foundation under Grant DMR-1752713.}

\acknowledgments{This  research  used  resources  of  the  National Energy  Research  Scientific  Computing  Center,  a  DOE Office of Science User Facility supported by the Office of Science of the U.S. Department of Energy under Contract No.  DE-AC02-05CH11231.}

\conflictsofinterest{The authors declare no conflict of interest.} 

\end{document}